\documentclass[epj, final]{svjour}
\usepackage{amsmath}
\usepackage{amssymb}
\usepackage[usenames,dvipsnames]{color}
\usepackage{graphics}
\begin{document}
\title{Asymmetric correlation matrices: an analysis of financial data}
\author{Giacomo Livan\inst{1} \and Luca Rebecchi\inst{2}
}                     
\offprints{glivan@ictp.it}          
\institute{Abdus Salam International Centre for Theoretical Physics, Strada Costiera 11, 34151 Trieste, Italy \and Dipartimento di Fisica Nucleare e Teorica, Universit\`a degli Studi di Pavia, Via Bassi 6, 27100 Pavia, Italy}
\date{Received: date / Revised version: date}
%
\abstract{
We analyze the spectral properties of correlation matrices between distinct statistical systems. Such matrices are intrinsically non symmetric, and lend themselves to extend the spectral analyses usually performed on standard Pearson correlation matrices to the realm of complex eigenvalues. We employ some recent random matrix theory results on the average eigenvalue density of this type of matrices to distinguish between noise and non trivial correlation structures, and we focus on financial data as a case study. Namely, we employ daily prices of stocks belonging to the American and British stock exchanges, and look for the emergence of correlations between two such markets in the eigenvalue spectrum of their non symmetric correlation matrix. We find several non trivial results, also when considering time-lagged correlations over short lags, and we corroborate our findings by additionally studying the asymmetric correlation matrix of the principal components of our datasets.
} 
\maketitle

\section{Introduction}
\label{intro}
A huge number of scientific disciplines, ranging from Physics to Economics, often need to deal with statistical systems described by a large number of degrees of freedom. Typically, it is very interesting, if not crucial, to analyze the correlations between the random variables describing such degrees of freedom. For this very reason, the development of both analytical and numerical tools to tackle the problem of correlation analysis is a fundamental topic in Multivariate Statistics. In most practical applications, one usually deals with a statistical system described in terms of $N$ random variables $\mathcal{R}_1, \ldots, \mathcal{R}_N$, and the most obvious thing to do in order to study such a system is to collect  as many observations as possible of such $\mathcal{R}_i$s. Then, assuming the $\mathcal{R}_i$s to be described by a stationary joint probability distribution, the observations can be used to compute empirical time averages of quantities expressed in terms of those variables. So, suppose $T$ equally spaced observations have been collected for each variable, and let us denote the time $t$ ($t = 1, \ldots, T$) observation of the random variable $\mathcal{R}_i$ ($i = 1,\ldots, N$) as $R_{it}$. Quite straightforwardly, one can collect all such numbers in a $N \times T$ matrix $\mathbf{R}$ whose generic entry reads $[\mathbf{R}]_{it} = R_{it}$. The most general correlation structure between the random variables $\mathcal{R}_i$ would read

\begin{equation} \label{gencorr}
\langle R_{it} R_{jt^\prime} \rangle = \mathcal{E}_{ij,tt^\prime}, 
\end{equation}
where $\langle \ldots \rangle$ denotes the expectation with respect to the joint probability density describing the $\mathcal{R}_i$s. However, in most practical applications the rather involved structure in equation \eqref{gencorr} can be factorized into its ``spatial'' and temporal parts. Assuming that the random variables $\mathcal{R}_i$s have zero mean and unit standard deviation, one could then write:

\begin{equation} \label{factcorr}
\langle R_{it} R_{jt^\prime} \rangle = C_{ij} \delta_{tt^\prime},
\end{equation}
and this will also be the case throughout the rest of this paper. In the previous expression, the matrix elements $C_{ij}$ (to be collected in a symmetric matrix $\mathbf{C}$) account for the cross-correlations amongst all possible pairs of variables in the system. On the other hand, the Kronecker delta in \eqref{factcorr} means that no auto-correlations are present in the system. Also, this means that each $C_{ij}$ in equation \eqref{factcorr} can be estimated as the following time average (where the data are assumed to be standardized):

\begin{equation} \label{Pearson}
c_{ij} = \frac{1}{T} \sum_{t=1}^T R_{it} R_{jt}.
\end{equation}
This expression is the very well-known Pearson estimator, and all the $c_{ij}$s can be collected in a $N \times N$ symmetric matrix

\begin{equation} \label{Pearsonmatrix}
\mathbf{c} = \frac{1}{T} \mathbf{R} \mathbf{R}^\mathrm{T},
\end{equation}
which represents a ``matrix estimator'' for the true correlation matrix $\mathbf{C}$ introduced in equation \eqref{gencorr}. So, the problem of characterizing the correlation structure of a statistical system essentially boils down to the estimation of the $N(N-1)/2$ independent entries of its correlation matrix from $NT$ empirical observations. However, depending on the length $T$ of the time series being used, the $C_{ij}$ estimates will inevitably be corrupted by a certain amount of measurement error, and this will eventually cause the whole correlation matrix $\mathbf{c}$ to be affected by the same problem. Several filtering recipes have been proposed in the statistical literature in order to partially clean correlation matrices from noise. On the other hand, a possible approach to attack the problem came from the Physics community, represented by the tools and methodologies developed in random matrix theory (RMT). Initially devised by Wigner \cite{Wigner} as a framework where to model the spectral properties of Hamiltionians of complex physical systems interacting through unknown laws, RMT gradually underwent a more formal evolution, eventually becoming a mathematical theory of its own \cite{Mehta,Anderson} and finding a plethora of application in extremely different scientific areas \cite{Handbook}. The main RMT result which is commonly used in correlation data analysis is the well known Mar\v cenko-Pastur distribution \cite{Marcenko}, \emph{i.e.} the average eigenvalue density for the correlation matrix of a system of uncorrelated Gaussian random variables in the ``thermodynamic limit'' $N, T \rightarrow \infty$, with $q \doteq T/N$ fixed. Such a distribution intuitively represents a suitable candidate for a ``null model'' with no correlations. Thus, any deviation between the Mar\v cenko-Pastur distribution and the empirically observed eigenvalue density of the data correlation matrix provides information about the correlation structure of the system under analysis. In the context of financial data analysis, this type of study was first carried out in the late nineties in \cite{Laloux99,Plerou99}, where the spectral properties of the correlation matrix of stocks belonging to the S$\&$P500 Index were analyzed over different time scales. Quite surprisingly, in those works most of the eigenvalue spectrum was shown to be fell fitted by a Mar\v cenko-Pastur distribution, whereas only few, larger, eigenvalues were shown to carry relevant information on the market correlation structure by ``leaking out'' of the Mar\v cenko-Pastur region. Ever since such works, physicists kept on analyzing financial correlation matrices, constantly refining the general picture described in \cite{Laloux99,Plerou99} with increasing levels of insight \cite{Mantegna99,Bonanno00,Laloux00,Plerou02,Guhr03,Bonanno04,Burda04,Raffaelli06,Drozdz07,Marsili09,Akemann10,Livan11}, and also generalizing the framework defined by equation \eqref{factcorr} to also include the effects due to temporal correlations \cite{Burda05,Burda10_bis}.

A quite natural generalization of the above picture is represented by the extension of correlation analyses to \emph{two} statistical systems $S_1$ and $S_2$, both described in terms of $N$ random variables. Then, one can straightforwardly write down the Pearson estimator \eqref{Pearson} for the correlation coefficient between the $i$th variable in $S_1$ and the $j$th variable in $S_2$:

\begin{equation} \label{Pearson12}
k_{ij} = \frac{1}{T} \sum_{t=1}^T R_{it}^{(1)} R_{jt}^{(2)}.
\end{equation}
Even more generally, one could think of the random variables in $S_1$ as a set of input variables, whose output is in turn described by the variables in $S_2$ (or vice versa). Then, it would be of great interest to further generalize \eqref{Pearson12} to the case of time lagged correlations, \emph{i.e.}

\begin{equation} \label{Pearson12tau}
k_{ij}(\tau) = \frac{1}{T-\tau} \sum_{t=1}^{T-\tau} R_{it}^{(1)} R_{j,t+\tau}^{(2)},
\end{equation}
so that equation \eqref{Pearson12} is recovered for $\tau = 0$. Recovering the previously outlined framework, it is of course convenient to collect all the $k_{ij}(\tau)$ estimates in a $N \times N$ matrix $\mathbf{k}(\tau)$. However, the most notable difference of such a matrix with respect to ``ordinary'' correlation matrices is that it is no longer symmetric, since $k_{ij}(\tau) \neq k_{ji}(\tau)$. Hence, its eigenvalues will in general be complex, and this feature, as we shall see later, will widely enrich the possible spectral analyses to be performed, and the subsequent considerations on the correlations between the two statistical systems to be studied.

In a financial context, it is quite interesting to interpret $S_1$ and $S_2$ as two different financial markets, so that the matrix $\mathbf{k}(\tau)$ will encode all of the relevant information on the possible correlations between them. In such a framework, we shall interpret $R_{it}^{(M)}$ as the standardized time $t$ log-return of the $i$th stock ($i = 1, \ldots, N$) in market $M$ ($M = 1,2$). Log-returns are the most commonly used variables in financial practice, and (at time $t$) they are defined as $\mathrm{log} \ S_{i,t}^{(M)} / S_{i,t-1}^{(M)}$, where $S_{i,t}^{(M)}$ denotes the time $t$ spot price of asset $i$ in market $M$.

The purpose of this paper is twofold. After briefly reviewing the most relevant spectral features of asymmetric correlation matrices as the one introduced in equation \eqref{Pearson12tau}, our first goal will be to look for an empirical realization of this type of matrices, providing some possible methodological guidelines to unravel the genuine correlations between two distinct complex systems. As anticipated, we choose financial data as a case study. So, our second main goal will be the one of verifying whether asymmetric correlation matrices can prove to be a valuable tool for the description of relevant stylized facts observed in financial markets. Admittedly, in this respect the choice of working with matrices of the type \eqref{Pearson12tau} represents a limitation, since one needs the matrix $\mathbf{k}(\tau)$ to be square (so it has eigenvalues), and this forces one to consider an equal number $N$ of stocks in the two markets. Working with singular values, as in \cite{Bouchaud07}, removes this constraint. However, we believe our first, more general, goal to justify such a limitation. 

Before we start to detail our study, it is worth mentioning that an analysis of financial data based on asymmetric matrices was first attempted in \cite{Drozdz06}. However, the random matrix benchmark used in that work was represented by the Ginibre orthogonal ensemble (GinOE), \emph{i.e.} the ensemble of random matrices with independent Gaussian real entries and no symmetry requirement. Despite producing complex eigenvalues, the spectral structure of the GinOE is completely different from the one produced by the random version of asymmetric correlation matrices as the one in equation \eqref{Pearson12tau}. Thus, we believe the analyses to be presented in our paper to be based on more solid theoretical grounds.

The paper is organized as follows. In Section \ref{Randomasym} the RMT results concerning the average eigenvalue density of random asymmetric correlation matrices will be overviewed. Then, the case study on financial data will be detailed in Section \ref{EmpAnal}, where the two-subsytems $S_1$ and $S_2$ will be represented by the American and British stock exchanges, respectively. The empirical results discussed in Section \ref{EmpAnal} will be corroborated in Section \ref{Joint} by investigating the spectral properties of the standard Pearson correlation matrix of the two datasets to be used. The paper will then be concluded with some final remarks in Section \ref{concl}.

\section{Random asymmetric correlation matrices}
\label{Randomasym}
The asymmetric correlation matrix in equation \eqref{Pearson12tau} can be clearly written as a product of two matrices:

\begin{equation} \label{Pearson12prod}
\mathbf{k}(\tau) = \frac{1}{T-\tau} \mathbf{R}_0^{(1)} (\mathbf{R}_\tau^{(2)})^\mathrm{T}
\end{equation}
where $[\mathbf{R}_l^{(1,2)}]_{it} = R^{(1,2)}_{i,t+l}$. In the following, we shall consider the case in which both matrices in the right hand side of equation \eqref{Pearson12prod} are random (in a sense to be made rigorous in a moment). Not many results are known on the spectra of products of random matrices (see for example \cite{Osborn04,Akemann05,Akemann09,Kanzieper10}) as the one in equation \eqref{Pearson12prod}, and most of them only describe ``microscopic'' spectral properties. However, in \cite{Burda10} an equation for the average eigenvalue density for a product of an arbitrary number of large Gaussian random matrices was derived. Such equation was derived by means of a planar diagram expansion (see \cite{Burda11} for a step by step introduction to this technique) under the assumption of all matrix dimensions going to infinity with their ratios kept fixed. Also, quite importantly for our present discussion, the aforementioned equation can be solved exactly for the product of two matrices, as in equation \eqref{Pearson12prod}. More precisely, assuming all matrix entries in both $\mathbf{R}^{(1)}_0$ and $\mathbf{R}^{(2)}_\tau$ to be independent and identically distributed Gaussian random numbers with zero mean and unit variance, the average eigenvalue density (in the complex plane) for the $\mathbf{k}^{(12)}$ matrix can be shown \cite{Burda10} to be:

\begin{equation} \label{Kdensity}
\rho_\mathbf{k} (\lambda,\lambda^*) = 
\left \{ \begin{array}{cc}
\frac{q^2}{\pi \sqrt{(1-q)^2 + 4q^2|\lambda|^2}} & \ \ \mathrm{for} \ \ |\lambda| \leq q^{-1/2} \\
0 & \ \ \mathrm{for} \ \ |\lambda| > q^{-1/2},
\end{array} \right.
\end{equation}
where again we have $q = T/N$ and $^*$ denotes complex conjugation. Thus, in the thermodynamic limit $N,T \rightarrow \infty$ with $q$ held fixed, the average eigenvalue density $\rho_\mathbf{k}$ displays circular symmetry within a circle of radius $q^{-1/2}$ centered in the origin of the complex plane. However, for any finite matrix dimension $N$, the circular symmetry is broken, due to the fact that $\mathrm{Tr}[\mathbf{k}^{(12)}(\tau)]$ is a real number, and this introduces a constraint on the eigenvalues. Thus, for any finite $N$ an excess of eigenvalues lying on the real axis, which can be shown to decrease	 as  
\begin{figure}
\begin{center}
\resizebox{0.95\columnwidth}{!}{
  \includegraphics{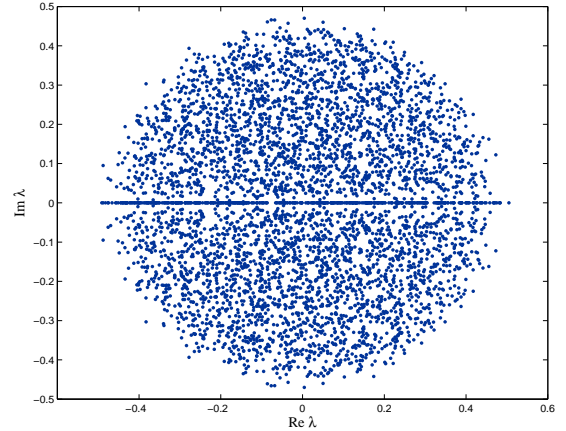}}
\end{center}
\caption{Eigenvalues of 50 random asymmetric correlation matrices with $N = 100$ and $T = 500$.}
\label{prod_eig}
\end{figure}
$\sqrt{N}$ \cite{Akemann07}, can be observed (see Figure \ref{prod_eig}). When considering complex rather than real entries for $\mathbf{k}^{(12)}$, circular symmetry is recovered also for finite values of $N$. Since the leading order (in $N$) results obtained for the eigenvalue densities with real and complex entries coincide, when taking the infinite matrix size limit one eventually ends up with the density in equation \eqref{Kdensity} in both cases. 

\begin{figure}
\begin{center}
\resizebox{0.95\columnwidth}{!}{
  \includegraphics{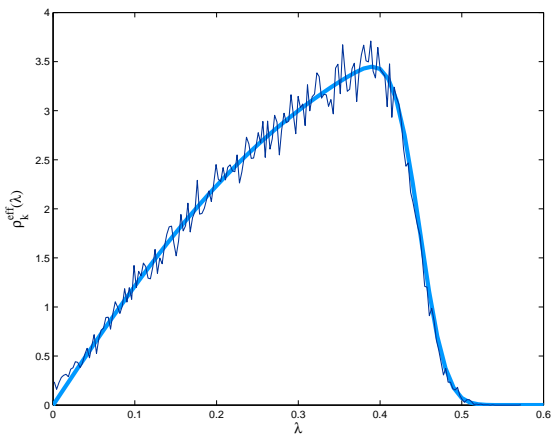}}
\end{center}
\caption{Radial density corresponding to the eigenvalues in Figure \ref{prod_eig} fitted with the effective finite size density of equation \eqref{Keffdensity} (finding $h = 27.9$).}
\label{rad_dens}
\end{figure}

Given the circular symmetry, one can safely work with the \emph{radial} eigenvalue density derived from \eqref{Kdensity}, which reads $\rho_\mathbf{k}^\mathrm{rad}(x) = 2 \pi x \rho_\mathbf{k}(\lambda,\lambda^*) |_{|\lambda| = x}$. Now, the thermodynamic limit density \eqref{Kdensity} reaches a finite value at the boundary of its domain ($|\lambda| = q^{-1/2}$), and then abruptly becomes equal to zero. However, when working with finite sized matrices, this transition is smoothed according to the following damping (conjectured in \cite{Burda10}, inspired by analogous finite size corrections that can be introduced rigorously for the Ginibre  random matrix ensembles \cite{Handbook,Forrester99}, and actually proved in \cite{Kanzieper10}):

\begin{equation} \label{Keffdensity}
\rho_\mathbf{k}^\mathrm{eff}(x) = \frac{1}{2} \rho_\mathbf{k}^\mathrm{rad}(x) \ \mathrm{erfc}(h(x-q^{-1/2})),
\end{equation}
where the parameter $h$ is phenomenological and needs to be adjusted by fitting. See Figure \ref{rad_dens}
for an example: as can be seen, the excess of eigenvalues on the real axis almost does not affect the overall shape of the radial density, even for relatively small matrix dimensions. Thus, in all of our following analyses we shall freely compare empirical data with the density in equation \eqref{Keffdensity}.

\section{Empirical analysis}
\label{EmpAnal}
In this section we shall look for an empirical realization of the asymmetric correlation matrix \eqref{Pearson12tau} in a financial context. Namely, as already anticipated, in the following we shall consider two different financial markets as the two statistical systems from which the data $R_{it}^{(1)}$ and $R_{j,t+\tau}^{(2)}$ (see again equation \eqref{Pearson12tau}) are drawn from. In particular, we shall focus on the American and British financial markets by employing prices of stocks belonging to the S$\&$P500 Index and the FTSE350 Index.

The dataset to be used is made of daily prices of $N = 200$ stocks (from both markets, so $400$ stocks overall) covering the years 2005-2011 ($T = 1595$ log-returns). It is important to remark here that, in order to empirically recreate the correlation matrix \eqref{Pearson12tau} (especially for $\tau = 0$, as in equation \eqref{Pearson12}) it is mandatory to work with data well defined on the same time steps $t = 1, \ldots, T$. For this very reason, prices collected from the American market during British holidays (and vice versa) were removed from the datasets.

When actually computing the eigenvalue spectrum of the generalized correlation matrix \eqref{Pearson12tau} for the aforementioned S$\&$P500 and FTSE350 datasets, two main features can be clearly distinguished: a main eigenvalue bulk close to zero and one large (in modulus) eigenvalue. We shall separately discuss those two aspects.

\subsection{The largest eigenvalue}
\label{LargestEig}
In the following, the variables $R^{(1)}_{it}$ in equation \eqref{Pearson12tau} will be meant to be the log-returns of stocks belonging to the S$\&$P500 Index, whereas the variables $R^{(2)}_{j,t+\tau}$ represent log-returns of stocks belonging to the FTSE350 Index. 

\begin{figure}
\begin{center}
\resizebox{0.95\columnwidth}{!}{
  \includegraphics{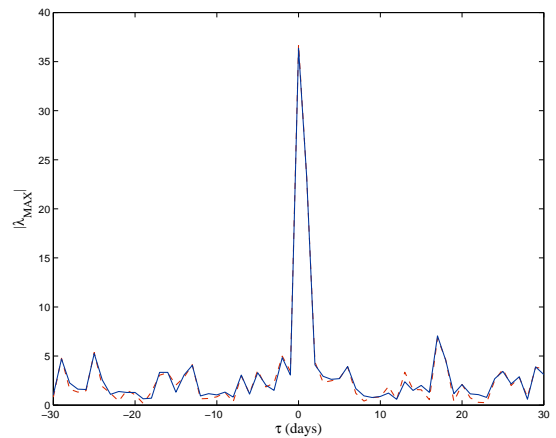}}
\end{center}
\caption{Absolute value of the largest eigenvalue $\lambda_\mathrm{MAX}$ of the asymmetric correlation matrix $\mathbf{k}(\tau)$ as a function of $\tau$.}
\label{LargestEigPlot}
\end{figure}

In Figure \ref{LargestEigPlot} the largest (in absolute value) eigenvalue $|\lambda_\mathrm{MAX}|$ is plotted as a function of $\tau$ (blue solid line). It is worth remarking that, except for a few cases, such eigenvalue is always found to be real. Intuitively, this is because it actually accounts for most of the trace of the $\mathbf{k}(\tau)$ matrix, which is a real number too. Now, as one can see from Figure \ref{LargestEigPlot}, the 
largest values of $|\lambda_\mathrm{MAX}|$ are found for $\tau = 0, 1$. More specifically, in both such cases $\lambda_\mathrm{MAX}$ is real and we have $\lambda_\mathrm{MAX}(\tau=0) = 36.4$ and $\lambda_\mathrm{MAX}(\tau=1) = 23.3$. Quite interestingly, one finds $\lambda_\mathrm{MAX}(\tau=-1) = 3.1$, much smaller than $\lambda_\mathrm{MAX}(\tau=1)$. This asymmetry highlights (also in the light of the interpretation of $\lambda_\mathrm{MAX}$ as average correlation to be discussed in the following) a strong influence of past American stock prices on the following day's British stock prices.

In order to verify the robustness of such evidence, we also computed the values of $\lambda_\mathrm{MAX}$ for $\tau = 0, \pm 1$ over eight different portions of our datasets (all of them made of 1195 daily log-returns and starting at $t = 1, 50, 100, \ldots, 350$). Over such eight samples we find, for $\tau = 0$, an average value of $\bar{\lambda}_\mathrm{MAX}(\tau = 0) = 37.3$ with a standard deviation $\sigma(\tau = 0) = 1.5$, whereas for $\tau = \pm 1$ we find $\bar{\lambda}_\mathrm{MAX}(\tau = 1) = 24.1$ and $\bar{\lambda}_\mathrm{MAX}(\tau = -1) = 3.8$ with standard deviations $\sigma(\tau = 1) = 0.1$ and $\sigma(\tau = -1) = 0.4$. In some cases, such estimates only appear to be close, but not perfectly compatible with the ones shown previously for the whole dataset. However, this fact does not point out any inconsistency, since the average and standard deviation values we reported are computed over (sometimes largely) overlapping time windows. So, they are not to be considered for any serious statistical comparison, and are only meant to qualitatively show how the estimates for $\lambda_\mathrm{MAX}$ fluctuate over time. 

For values of $\tau$ other than $0$ and $\pm 1$, $|\lambda_\mathrm{MAX}|$ seems to follow a random path, approximately lying between 0 and 10. The interesting point, however, is that $\lambda_\mathrm{MAX}$ is very often found to be much larger than the limiting radius predicted by RMT for the eigenvalue density of random asymmetric correlation matrices. As already detailed in the previous section, such a radius is equal to $q^{-1/2} = \sqrt{N/T}$ (see equation \eqref{Kdensity}). With the values of $N$ and $T$ of our dataset we have $R \sim 0.35$, much smaller than most values of $|\lambda_\mathrm{MAX}|$. At first, this might seem to suggest the existence of some non trivial long-range correlation. On the contrary, such persistently high values can be shown to be a spurious effect by means of the following argument. Let $\bar{k}(\tau)$ be the average estimated correlation between stocks in the two markets, \emph{i.e.}

\begin{equation} \label{meank}
\bar{k}(\tau) = \frac{1}{N^2} \sum_{i,j=1}^N k_{ij}(\tau)
\end{equation}
with $k_{ij}(\tau)$ defined as in equation \eqref{Pearson12tau}. Let us then approximate the whole matrix as $\mathbf{k}(\tau) \sim \bar{k}(\tau) \mathbf{E}_N$, where $\mathbf{E}_N$ is the $N \times N$ matrix whose entries are all equal to one: this amounts to approximate all correlations in $\mathbf{k}(\tau)$ with their average. Now, it can be easily shown that the matrix $\mathbf{E}_N$ has one eigenvalue equal to $N$ and $N-1$ eigenvalues equal to zero. Under such a ``mean field'' approximation the eigenvalue spectrum of $\mathbf{k}(\tau)$ would read

\begin{eqnarray} \label{approxspectrum}
\mathrm{det} (\mathbf{k}(\tau) - \lambda \mathbf{1}_N) &\sim& \mathrm{det} (\bar{k}(\tau) \mathbf{E}_N - \lambda \mathbf{1}_N) \\ \nonumber &=& (- \lambda)^{N-1} (\bar{k}(\tau) N - \lambda),
\end{eqnarray}
where $\mathbf{1}_N$ represents the $N \times N$ identity matrix. Equation \eqref{approxspectrum} means that we would have $N-1$ zero modes plus one eigenvalue equal to $\bar{k}(\tau) N$. Quite remarkably, this simple and apparently very rough approximation is actually enough to explain the persistence of a large eigenvalue over large time lags: the red dashed line in Figure \ref{LargestEigPlot} represents $|\bar{k}(\tau)|N$, and one can see how close this follows the path of the largest eigenvalue $|\lambda_\mathrm{MAX}|$. All in all, this latter merely reflects the average correlation for a certain value of the time lag $\tau$. Most importantly, this is also true for $\tau = 0,1$, i.e. when $|\lambda_\mathrm{MAX}|$ reaches its highest measured values, and such evidence tells us, unsurprisingly, that the average correlations are much higher for those values of $\tau$. For other values of $\tau$, the absolute value of the average correlation approximately lies between 0 and $\pm 0.04$ (\emph{i.e.} very small values), but the enhancing factor $N$ causes the corresponding large eigenvalue $|\bar{k}(\tau)|N$ to lie between 0 and 10, as already stated. In the genuinely random matrix model for $\mathbf{k}(\tau)$ outlined in Section \ref{Randomasym} the average correlation $\bar{k}(\tau)$ is very strongly suppressed, so that no large and isolated eigenvalues can appear. 

\subsection{Bulk of the spectrum}
\label{BulkSpectrum}
\begin{figure*}
\begin{center}
\resizebox{0.95\columnwidth}{!}{
  \includegraphics{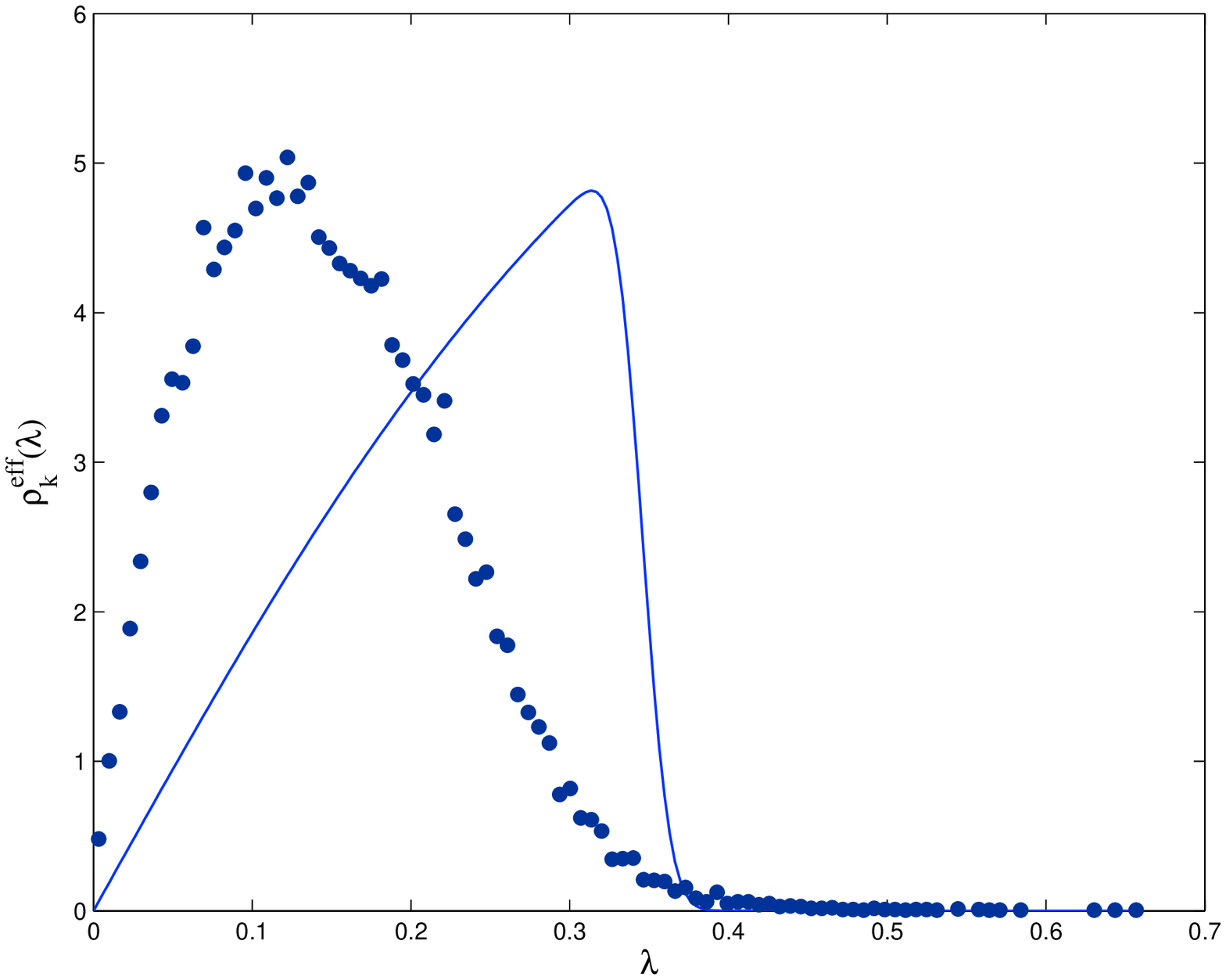}}
\hspace{1cm}
\resizebox{0.95\columnwidth}{!}{
  \includegraphics{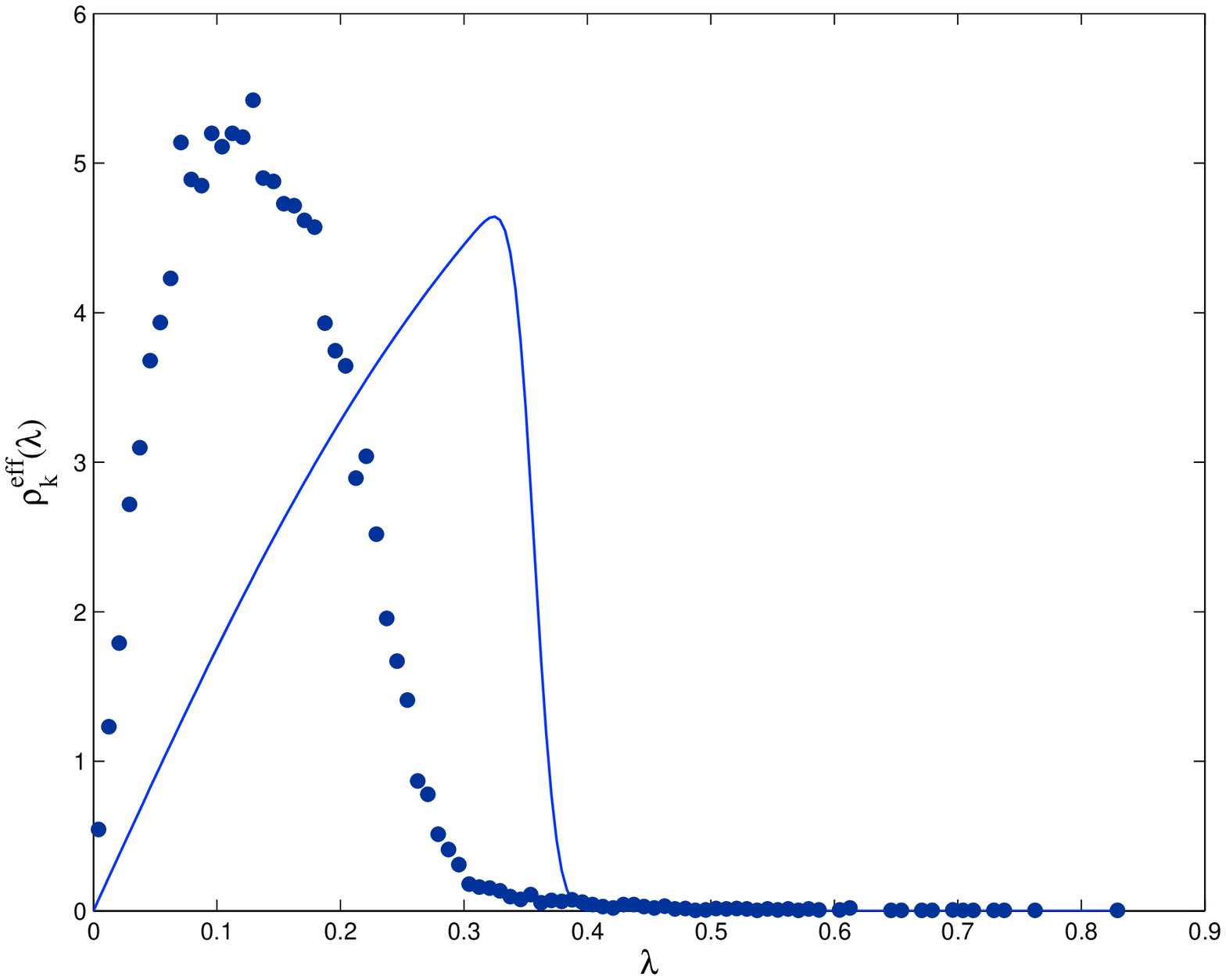}}
\end{center}
\caption{Eigenvalue spectrum of the asymmetric correlation matrix $\mathbf{k}(\tau)$ for stocks belonging to the S$\&$P500 and FTSE350 indices (blue dots), with $N = 190$, $T = 1595$ and $\tau = 0$ (left), $T = 1495$ and $\tau = 100$ (right). The spectrum statistics is enhanced by means of a bootstrap approach (see main text). The solid line represents the effective radial density predicted by RMT (equation \eqref{Keffdensity}) for the same values of $N$ and $T$ . The $h$ parameter was adjusted by fitting on a Monte Carlo density with large statistics, yielding $h = 51.80$ and $h = 50.86$ in the two cases.}
\label{Bulk}
\end{figure*}

In Figure \ref{Bulk} the main part of the radial eigenvalue spectrum of the the $\mathbf{k}(\tau)$ matrix constructed with the aforementioned S$\&$P and FTSE datasets is plotted (blue dots) for $\tau = 0$ and
$\tau = 100$. In order to improve the statistics, a bootstrap approach is followed: namely, 200 iterations are performed and, for each of those, the $\mathbf{k}(\tau)$ matrix is constructed by randomly selecting 190 stocks out of the 200 available ones for each of the two stock sets. This is done under the reasonable assumption that the eigenvalue spectrum will not be drastically affected, at least in its overall appearance, by the particular stock selection. Also, in both plots of Figure \ref{Bulk}, the effective radial eigenvalue density \eqref{Keffdensity} predicted by RMT for $N = 190$ and $T = 1595$ ($T = 1495$ for the case $\tau = 100$) is shown. In both cases the $h$ parameter was determined by fitting on Monte Carlo densities with very large statistics. As is immediate to see, for both of the considered values of $\tau$ the empirical and theoretical densities have no similarity at all. Also, trying to fit the effective density \eqref{Keffdensity} allowing the $q$ ratio to be a free parameter (much in the same spirit of what was done in \cite{Laloux99,Plerou99} with the Mar\v cenko-Pastur density) does not provide acceptable results, essentially due to to the much slower falloff of the empirical densities with respect to the exponential one of the RMT radial density \eqref{Keffdensity}. At first, one might naively interpret such discrepancies, especially the one for $\tau = 100$, as a sign of some long-range time correlations between the markets under study. However, one should recall that the RMT densities in equations \eqref{Kdensity} and \eqref{Keffdensity} are derived for the $\mathbf{k}(\tau)$ matrix in \eqref{Pearson12tau} under the assumption that the two sub-systems have no mutual correlation and no correlation \emph{of their own}. This is a crucial point: using a large time lag $\tau$ should suppress all correlations between the stocks in the two datasets, and this is actually confirmed by the previous analysis on the $\tau$-dependence of the average correlation $\bar{k}$. However, using a sliding time window does not suppress the self-correlations within \emph{each} market: figuratively speaking, those are ``dragged along'' by the sliding window $\tau$ itself. Hence, one should try to disentangle the two different types of correlations, getting rid of the inner ones while retaining only those existing between the two sub-systems. Quite naturally, this task can be accomplished by mapping the original variables onto the corresponding sets of principal components.

\subsection{Mapping onto principal components}
\label{PCmap}
Starting from our two datasets, let us construct their standard Pearson correlation matrices in the usual way as $\mathbf{C}^{(1)} = \mathbf{R}^{(1)} (\mathbf{R}^{(1)})^\mathrm{T} / T$ and $\mathbf{C}^{(2)} = \mathbf{R}^{(2)} (\mathbf{R}^{2})^\mathrm{T} / T$. Denoting their eigenvalues as $\lambda_{1,i}$ and $\lambda_{2,j}$ (for $i,j = 1, \ldots, N$), and the corresponding eigenvectors as $\mathbf{V}^{(1,i)} = (V_1^{(1,i)}, \ldots, V_N^{(1,i)})$ and $\mathbf{V}^{(2,j)} = (V_1^{(2,j)}, \ldots, V_N^{(2,j)})$, principal components are defined as follows:

\begin{equation} \label{PC}
e_{it}^{(M)} = \frac{1}{\sqrt{\lambda_{M,i}}} \sum_{j=1}^N V_j^{(M,i)} R_{jt}^{(M)},
\end{equation}
where $M = 1,2$ and $R_{jt}^{(1)}$ and $R_{jt}^{(2)}$ are as in equation \eqref{Pearson12tau}. Now, exploiting eigenvector orthogonality one can immediately verify that principal components are \emph{exactly uncorrelated}:

\begin{equation} \label{PCuncorr}
\frac{1}{T} \sum_{t=1}^T e_{it}^{(M)} e_{jt}^{(M)} = \delta_{ij}.
\end{equation}
Moreover, inverting equation \eqref{PC} one can expand any of the original variables in terms of principal components:

\begin{equation} \label{PCdec}
R_{it}^{(M)} = \sum_{j=1}^N \sqrt{\lambda_{M,j}} \ V_i^{(M,j)} e_{jt}^{(M)}.
\end{equation}
This relation is exact and shows that any of the random variables $R_{it}^{(M)}$ can be decomposed over a set of uncorrelated variables, whose explanatory power (in terms of variance) of the original variables' dynamics can be ranked depending on the size of the corresponding eigenvalues.

\begin{figure}
\begin{center}
\resizebox{0.99\columnwidth}{!}{
  \includegraphics{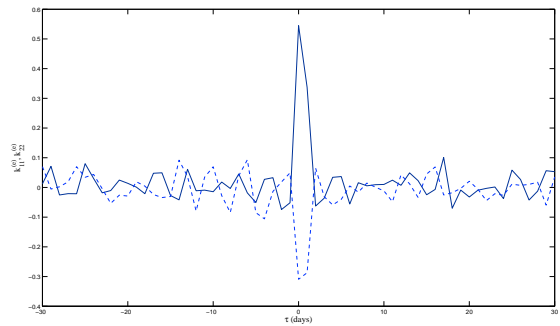}}
\end{center}
\caption{Correlation coefficients $k_{11}^{(e)}(\tau)$ (solid line) and $k_{22}^{(e)}(\tau)$ (dashed line).}
\label{K11}
\end{figure}

\begin{figure}
\begin{center}
\resizebox{0.99\columnwidth}{!}{
  \includegraphics{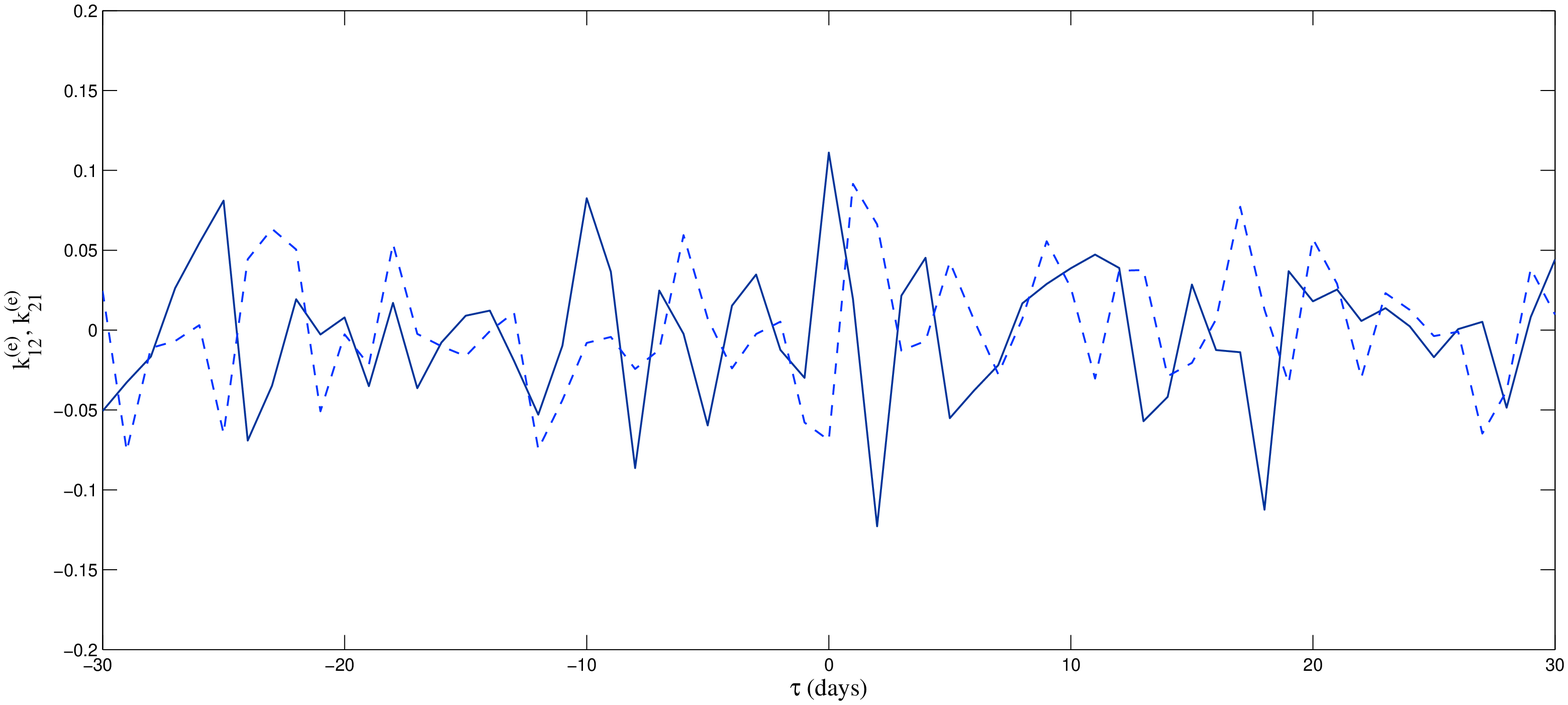}}
\end{center}
\caption{Correlation coefficients $k_{12}^{(e)}(\tau)$ (solid line) and $k_{21}^{(e)}(\tau)$ (dashed line).}
\label{K12}
\end{figure}

Principal components look as a quite appealing set of variables to use in the framework of asymmetric correlation matrices between two distinct financial markets. As already stated, the huge and persistent (over large time lags) deviations between empirical spectra and RMT predictions seem to be due to the inner correlations of the two markets. Switching to principal components circumvents this problem. Let us then introduce the asymmetric correlation matrix between principal components of the two datasets in use. We shall write the correlation coefficients as

\begin{equation} \label{corrPC}
k_{ij}^{(e)}(\tau) = \frac{1}{T-\tau} \sum_{t=1}^{T-\tau} e_{it}^{(1)} e_{j,t+\tau}^{(2)},
\end{equation}
and we shall collect them in a matrix $\mathbf{k}^{(e)}(\tau)$. So now, since the principal components in each set are completely uncorrelated, any deviation of the $\mathbf{k}^{(e)}(\tau)$ matrix's eigenvalue spectrum from the pure noise RMT prediction can only be imputed to correlations between the two sub-systems under study, encoded as correlations between their respective principal components. Even more interestingly, as is quite well known, the first few principal components, \emph{i.e.} the dominant ones related to the largest eigenvalues, can be given a simple financial interpretation (see for example \cite{Plerou02}): the first one arises as a consequence of collective market fluctuation (hence it is usually given the name of ``market mode''), and the first few after that generally correspond to market sectors. Hence, before studying the whole spectrum of the $\mathbf{k}^{(e)}(\tau)$ matrix, let us take a look at the correlations between such variables. In Figures \ref{K11} and \ref{K12} the 
$\tau$-dependence of some matrix elements in $\mathbf{k}^{(e)}(\tau)$ is shown. Namely, in Figure \ref{K11} the correlation coefficients $k_{11}^{(e)}$ and $k_{22}^{(e)}$ between the two main principal components (\emph{i.e.} those related to the two largest eigenvalues) in the two datasets is plotted. Such principal components account for $46.5\%$ of the overall data variance in the S$\&$P dataset, and for $31.7\%$ in the FTSE dataset. As one can see, quite strong correlations (either positive or negative) are again found for $\tau = 0,1$: $k_{11}(\tau=0) = 0.54$, $k_{11}(\tau=1) = 0.34$ and $k_{22}(\tau=0) = -0.31$, $k_{22}(\tau=1) = -0.29$. For different values of $\tau$, much smaller values are found, similarly to the case of the largest eigenvalue (see Figure \ref{LargestEigPlot}). On the contrary, correlations between the first and second principal components in the two datasets, encoded in the matrix elements $k_{12}^{(e)}$ and $k_{21}^{(e)}$are found to be quite small for all values of $\tau$ (see Figure \ref{K12}). Similar facts, \emph{i.e.} strong ``diagonal'' correlations for $\tau = 0,1$ and weak ``off-diagonal'' correlations, are observed also when considering the other most relevant principal components. This is quite interesting since, apart from the first component $e_1^{(1)}$ and $e_1^{(2)}$, which represent market modes, the other most relevant principal components do not necessarily represent one well-defined market sector or the same sector in the two markets. Nevertheless, their quite strong mutual correlations for $\tau = 0,1$ suggest that they encode relevant information about ``orthogonal'' (in the sense made rigorous by PCA) market portions, which remain ``orthogonal'' across different financial markets (as demonstrated by the small ``off-diagonal'' correlations $k_{ij}^{(e)}(\tau)$ for $i \neq j$). 

As a concluding remark to this discussion, let us also clarify how the correlations between different principal components impact those between the ``true'', original variables (daily log-returns in our case). Starting from the correlation coefficient \eqref{Pearson12tau}, and using equation \eqref{PCdec}, one finds

\begin{eqnarray} \label{corrlink}
k_{ij}(\tau) &=& \frac{1}{T-\tau} \sum_{t=1}^{T-\tau} R_{it}^{(1)} R_{j,t+\tau}^{(2)} \\ \nonumber
&=& \frac{1}{T-\tau} \sum_{t=1}^{T-\tau} \sum_{l,s=1}^1 \sqrt{\lambda_{1,l} \lambda_{2,s}} \ V_i^{(1,l)} V_j^{(2,s)} e_{lt}^{(1)} e_{s,t+\tau}^{(2)} \\ \nonumber
&=& \sum_{l,s=1}^N \sqrt{\lambda_{1,l} \lambda_{2,s}} \ V_i^{(1,l)} V_j^{(2,s)} k_{ls}^{(e)}(\tau),
\end{eqnarray}
and from this relation one sees that, unsurprisingly, the largest eigenvalues and largest correlations between principal components justify, for most part, the correlations between the original variables. Defining two $N \times N$ matrices $\mathbf{W}^{(1)}$ and $\mathbf{W}^{(2)}$ with entries

\begin{equation} \label{Wmatr}
W_{ij}^{(M)} = \sqrt{\lambda_{M,j}} \ V_i^{(M,j)}
\end{equation}
(where $M = 1,2$) allows us to rewrite equation \eqref{corrlink} in matrix form:

\begin{equation} \label{corrlink2}
\mathbf{k}(\tau) = \mathbf{W}^{(1)} \mathbf{k}^{(e)}(\tau) \left ( \mathbf{W}^{(2)} \right )^\mathrm{T}.
\end{equation}
We shall come back later to this point.

\begin{figure*}
\begin{center}
\resizebox{0.95\columnwidth}{!}{
  \includegraphics{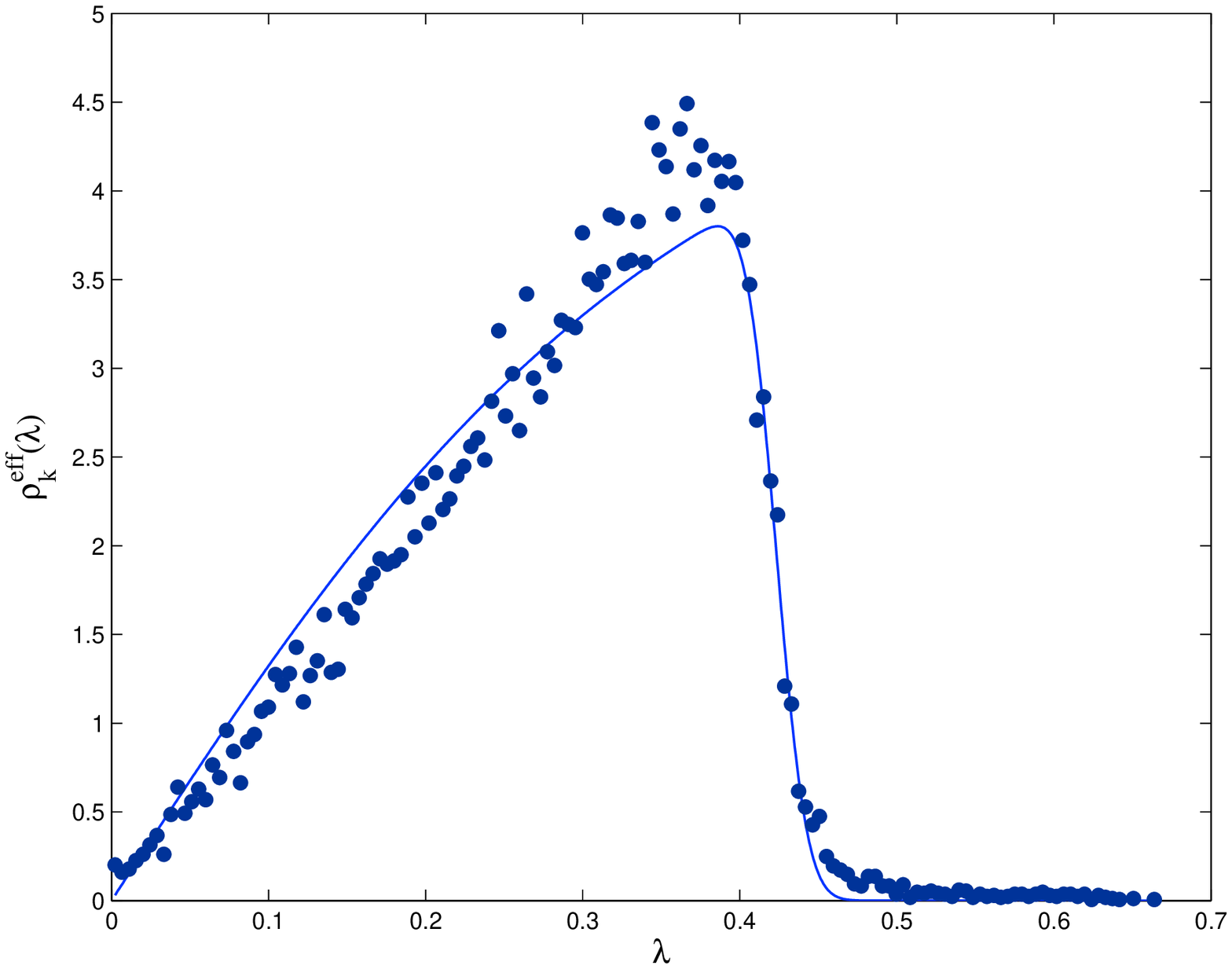}}
\hspace{1cm}
\resizebox{0.95\columnwidth}{!}{
  \includegraphics{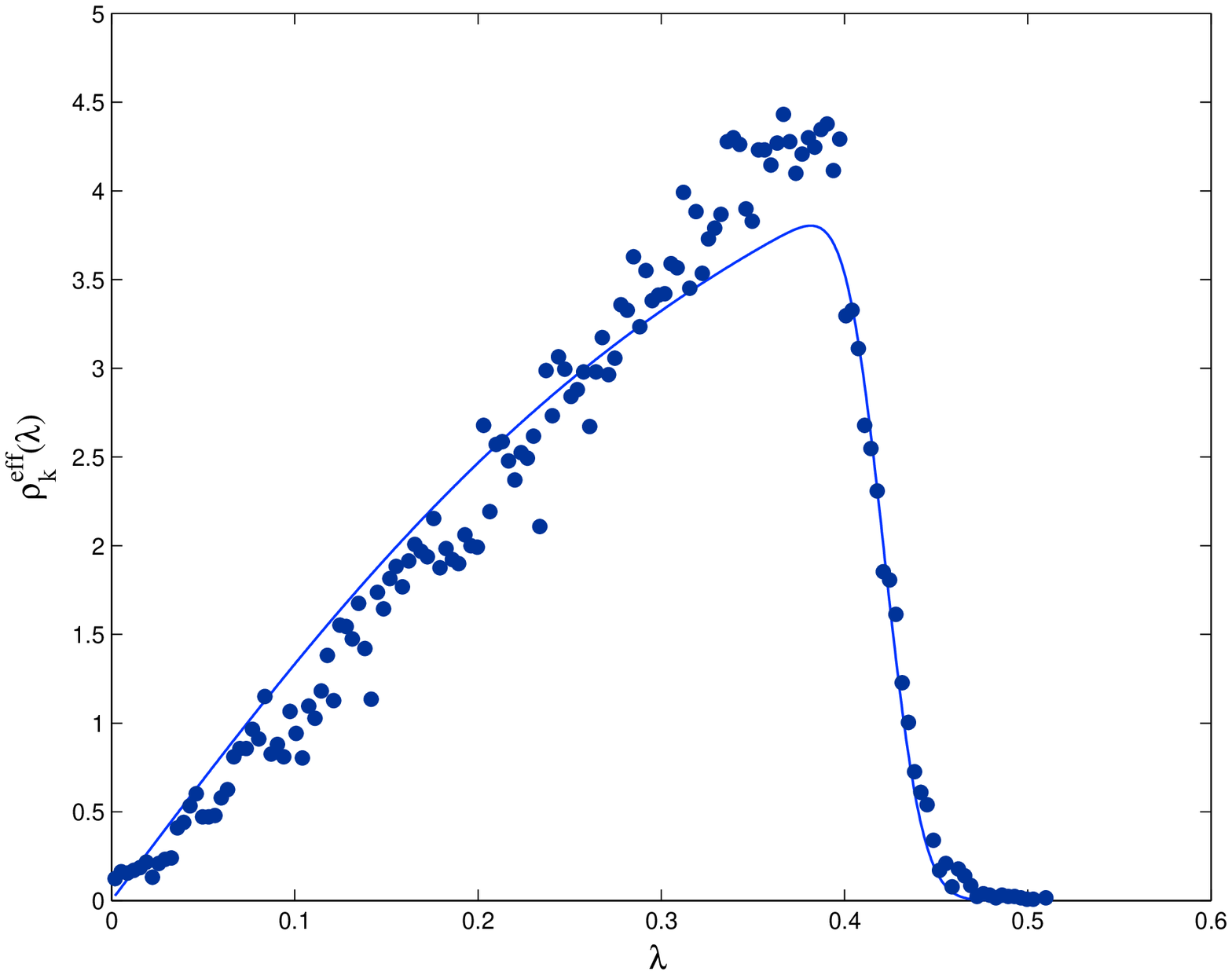}}
 \hspace{1cm}
 \resizebox{0.95\columnwidth}{!}{
  \includegraphics{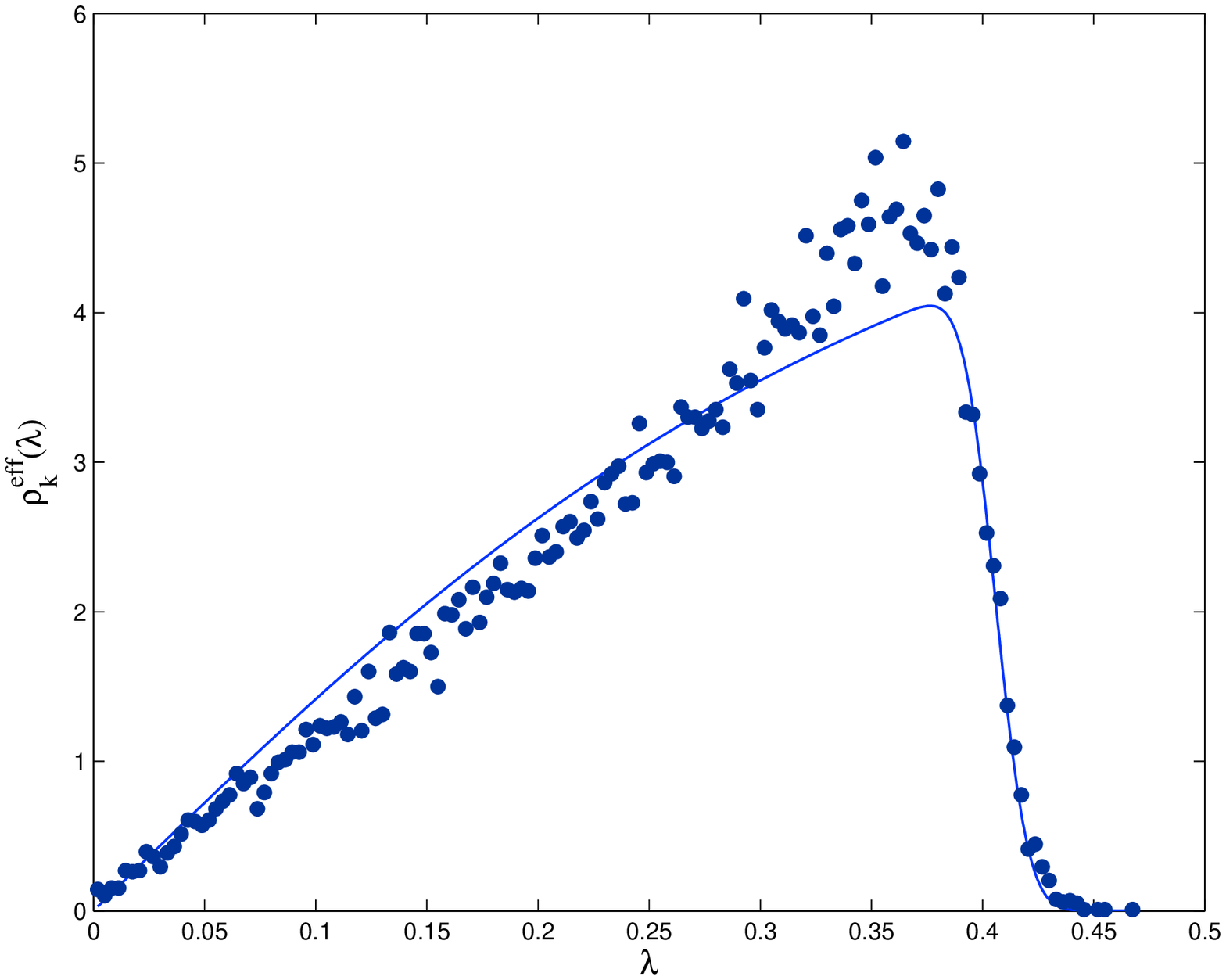}} 
\end{center}
\caption{Empirical eigenvalue spectra (enhanced by bootstrap) of the $\mathbf{k}^{(e)}(\tau)$ correlation matrix \eqref{corrPC} of principal components for $\tau = 0$ (top-left), $\tau = 1$ (top-right) and $\tau = 30$ (bottom) fitted with the radial density \eqref{Keffdensity} (solid line).}
\label{PCHistPlot}
\end{figure*}

\begin{figure*}
\begin{center}
\resizebox{0.95\columnwidth}{!}{
  \includegraphics{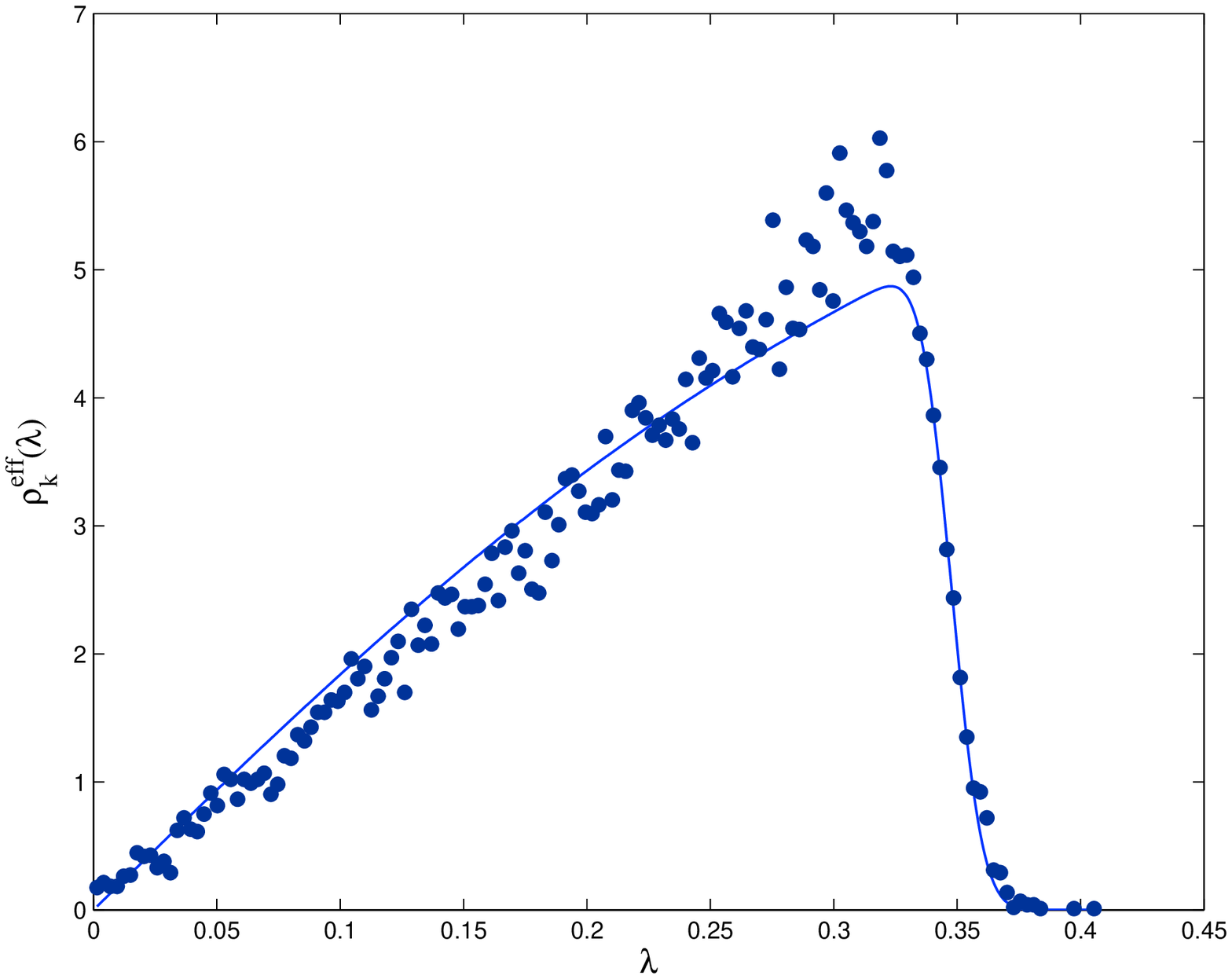}}
\hspace{1cm}
\resizebox{0.95\columnwidth}{!}{
  \includegraphics{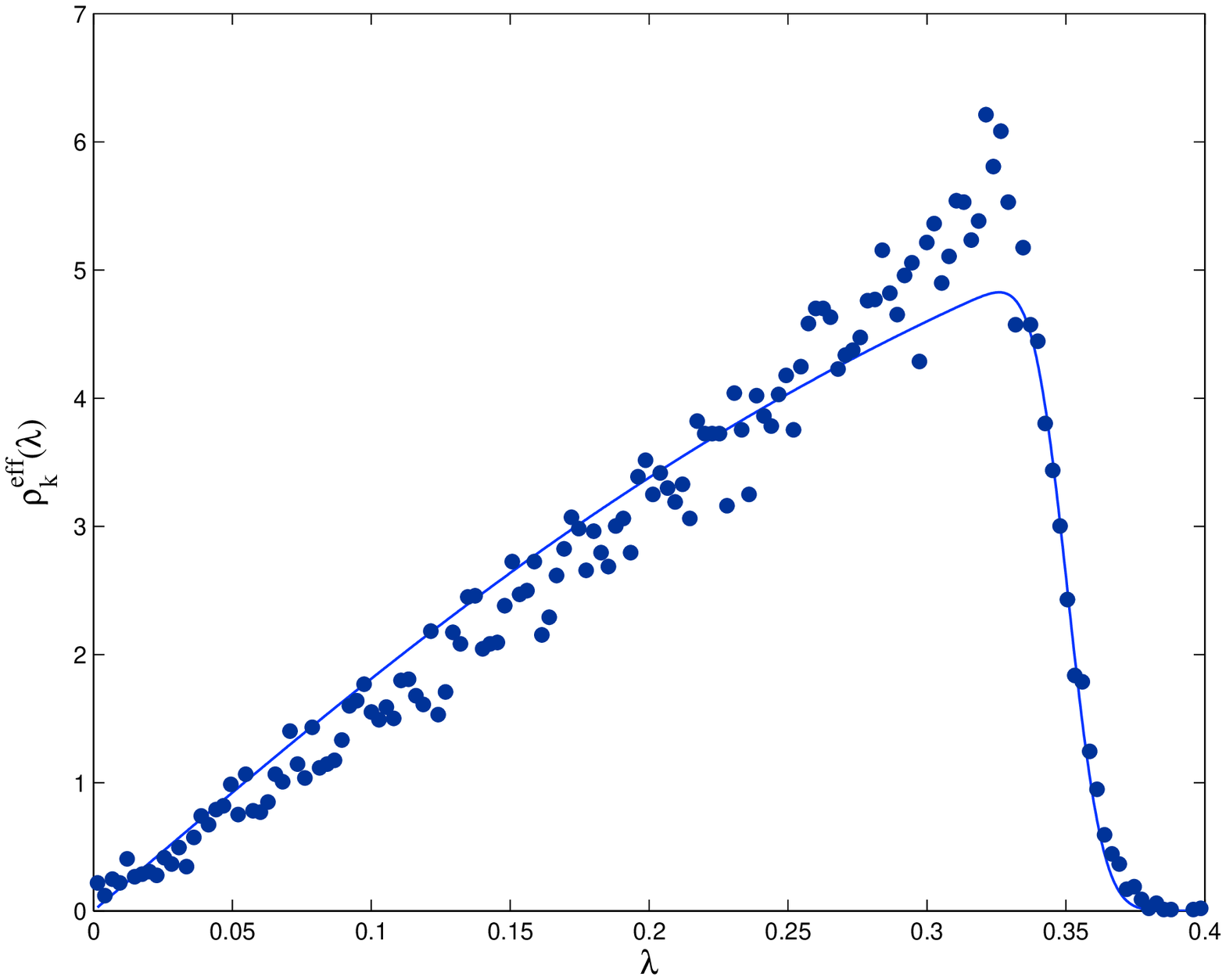}}
\end{center}
\caption{Empirical eigenvalue density of the asymmetric correlation matrix of principal components (equation \eqref{corrPC}) after performing one same time reshuffling on all the stocks in the S$\&$P dataset and another one on all stocks in the FTSE dataset (see the main text for more details on this). The left plot refers to the case $\tau = 0$, while in the right plot we have $\tau = 30$.}
\label{PCHistResh}
\end{figure*}

Finally, let us look at the eigenvalue spectrum of the asymmetric correlation matrix $\mathbf{k}^{(e)}(\tau)$ of the principal components. In Figure \ref{PCHistPlot} the empirical radial eigenvalue spectra of the $\mathbf{k}^{(e)}(\tau)$ matrix are plotted for $\tau = 0,1,30$ (top-left, top-right and bottom, respectively). In all cases, the 
same bootstrap approach already adopted for the spectra in Figure \ref{Bulk} is used, \emph{i.e.} 200 iterations are performed, each time randomly selecting $N = 190$ stocks out of the 200 available ones in each dataset. As can be seen, for all values of $\tau$ one now ends up with an eigenvalue spectrum which is much closer to the one predicted by RMT (solid line in all plots of Figure \ref{PCHistPlot}) than when using the original variables (Figure 	\ref{Bulk}). For $\tau = 0,1$ significant correlations between the two markets under study exist, as pointed out in the previous analyses, and this is reflected into visible deviations between the empirical and the theoretically expected eigenvalue density. For larger values of $\tau$ (exemplified by the bottom plot of Figure \ref{PCHistPlot}) the overall agreement improves: the exponential falloff of the RMT density is quite well reproduced (whereas for $\tau = 0,1$ this is not the case), but still an excess of eigenvalues lying around the peak region of the distribution can be clearly seen. However, even though the agreement between data and theory is still not excellent even after switching to principal components, the main point to be discussed is the following: namely, all the theoretical densities in Figure \ref{PCHistPlot} are fitted to the empirical histograms, allowing \emph{both} \emph{h} and \emph{q} (see equation \eqref{Keffdensity}) to be free parameters. Now, whereas the former parameter is phenomenological by definition, the latter should in principle be given by the ratio $T/N$. The values of $N$ and $T$ used in Figure \ref{PCHistPlot} give $q \sim 8.4$ while by fitting one obtains $q = 5.59$, $q = 5.64$ and $q = 6.08$ for $\tau = 0,1,30$ respectively: in all cases the effective $q$ parameter is very different with respect to its expected value. Moreover, one can also check that by performing \emph{one same} time reshuffling for all the $e^{(1)}$s and another one (different from the first) for all the $e^{(2)}$s, the expected value of is essentially reached (see Figure \ref{PCHistResh}, where the radial density \eqref{Keffdensity} is fitted giving $q = 8.24$ 
when $\tau = 0$ and $q = 8.15$ when $\tau = 30$, very close to the ``natural'' value $q \sim 8.4$. So, how to interpret this result?

Performing one time reshuffling within one dataset and a different one within the other one has the following effects on the different types of correlations involved:
\begin{itemize}
	\item Performing one same reshuffling for all the variables within one set keeps their mutual cross-correlations intact. Since the variables being dealt with here are principal components, this kind of reshuffling keeps them uncorrelated (see equation \eqref{PCuncorr}).
	\item Since the two reshufflings performed on the two datasets are different, all correlations between variables belonging to different sub-systems are destroyed.
	\item Performing a time reshuffling on a time series reasonably destroys all possible autocorrelations in it.
\end{itemize}
As a matter of fact the first two points in the above list empirically recreate the conditions under which the RMT density \eqref{Keffdensity} is derived, \emph{i.e.} no self-correlations within each system and no correlation between the two. However, such conditions are essentially obtained also when the correlations amongst principal components are computed for large enough values of $\tau$ (see Figures \ref{K11} and \ref{K12}), whereas the example shown in Figure \ref{PCHistPlot} shows that this is not the case, since one ends up with an effective value of $q$ which is quite 
\begin{figure*}
\begin{center}
\resizebox{0.95\columnwidth}{!}{
  \includegraphics{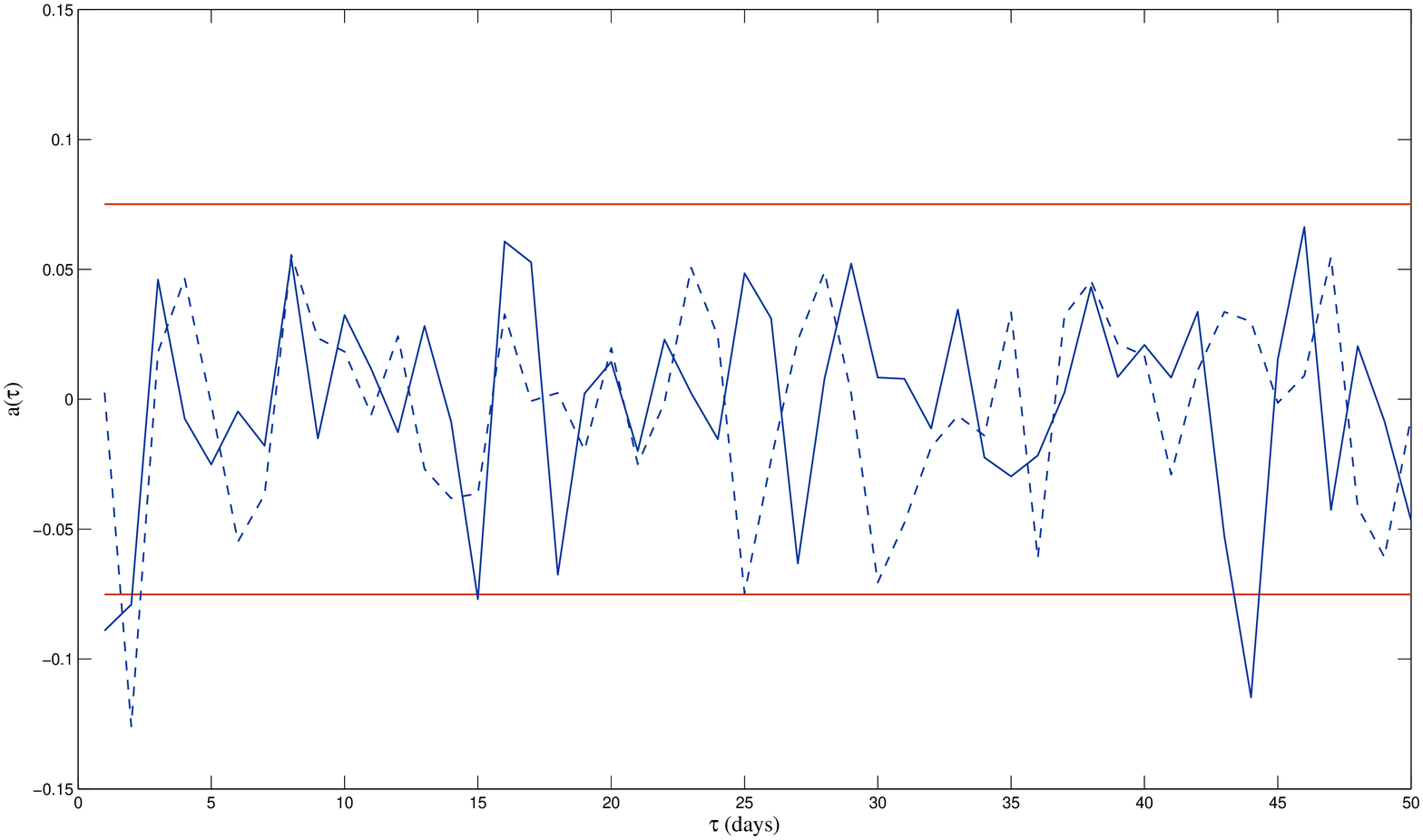}}
\hspace{1cm}
\resizebox{0.95\columnwidth}{!}{
  \includegraphics{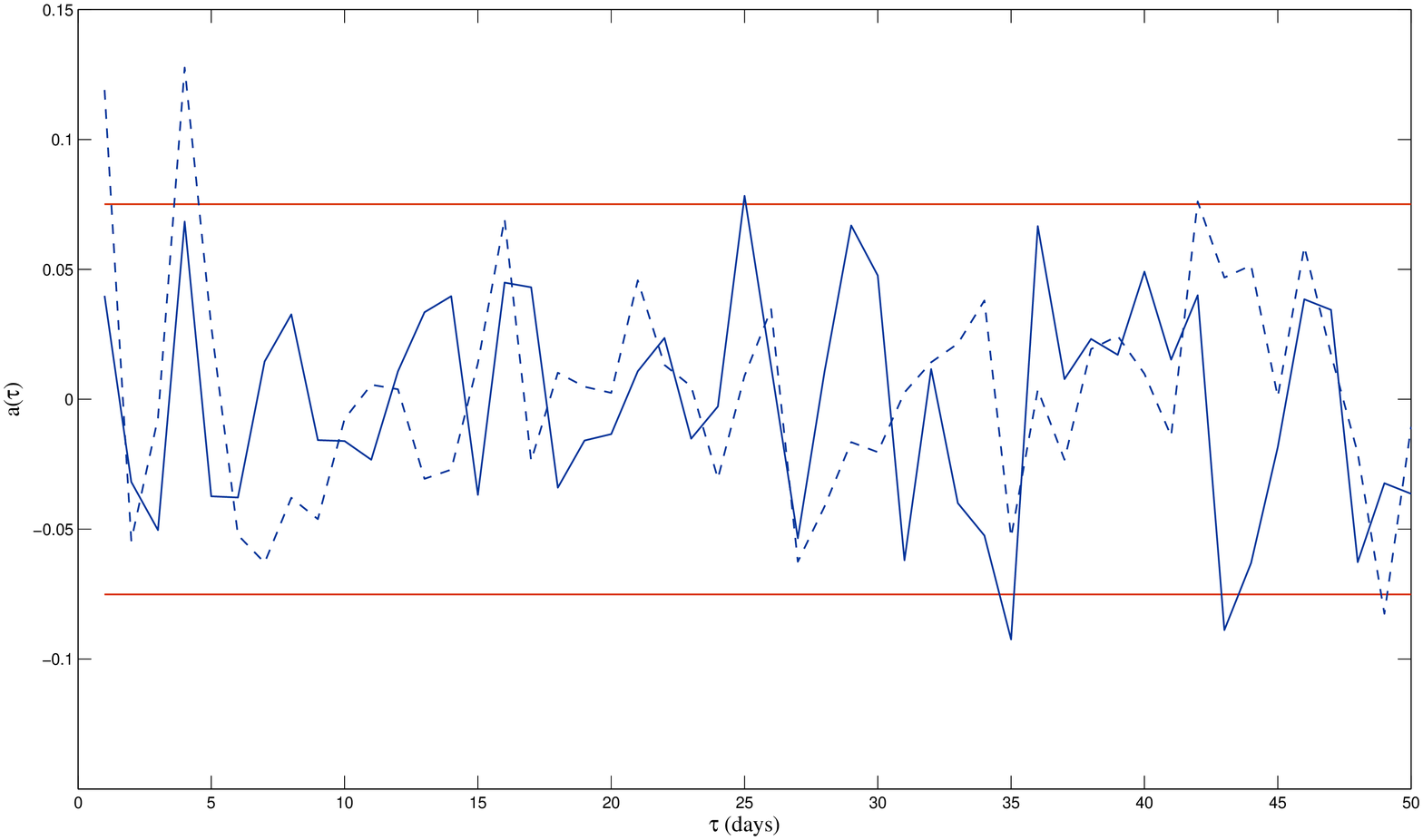}}
\end{center}
\caption{Autocorrelation function of the first two principal components of the S$\&$P (left plot) and FTSE (right plot) datasets. In both plots, solid lines refer to the first principal component, while dashed lines refer to the second one. The region delimited by the red horizontal lines represents the $99.7\%$ confidence interval for the autocorrelation of a purely random process.}
\label{PCAcorr}
\end{figure*}
far from the expected one. So, the last point in the above list appears to be the crucial one. 

Finding smaller values of $q$, with respect to the ``natural'' ones, as in Figure \ref{PCHistPlot}, amounts to larger effective values of $N$ or smaller effective values of $T$, and the latter seems the only possible option. As a matter of fact, principal component analysis grants us that no cross-correlations exist between principal components (equation \eqref{PCuncorr}). Nevertheless, nothing prevents such variables to display autocorrelations, contrarily to the original variables, \emph{i.e.} the log-returns, which are known not to display any relevant autocorrelation (see for example \cite{Campbell97}). In this respect, see Figure \ref{PCAcorr}, where the autocorrelations (as a function of $\tau$) of the first two principal components for each of our datasets are plotted; the $99.7\%$ confidence interval for a purely random process of length $T$ is shown in red. As can be clearly seen, the interval boundaries are crossed several times, thus illustrating that, indeed, the main principal components do feature autocorrelations (similar behaviors are found for all other principal components). On a qualitative level, the presence of autocorrelations reduces the number of degrees of freedom in the system, and justifies the need to accordingly adjust $T$ to an effective dimensionality \cite{Bretherton90}.

\section{Joint correlation matrix}
\label{Joint}
Before concluding, let us complement our analyses on asymmetric correlation matrices by studying the standard correlation matrix of our whole dataset. Let us then consider the following $2N \times T$ matrix:

\begin{equation} \label{bigmatr}
\mathbf{R} = \left ( \begin{array}{c}
	\mathbf{R}^{(1)} \\
	\mathbf{R}^{(2)}
	\end{array} \right ),
\end{equation}
where $\mathbf{R}^{(1)}$ and $\mathbf{R}^{(2)}$ are two $N \times T$ matrices containing the time series of our S$\&$P and FTSE datasets, respectively. From the matrix in equation \eqref{bigmatr}, we can build the ordinary Pearson correlation matrix as in equation \eqref{Pearsonmatrix}:

\begin{equation} \label{jointmatr}
\mathbf{c} = \frac{1}{T} \mathbf{R} \mathbf{R}^\mathrm{T} =
\left ( \begin{array}{cc}
	\frac{\mathbf{R}^{(1)} \left (\mathbf{R}^{(1)} \right)^\mathrm{T}}{T} & \frac{\mathbf{R}^{(1)} \left ( \mathbf{R}^{(2)} \right )^\mathrm{T}}{T}  \\
	\frac{\mathbf{R}^{(2)} \left ( \mathbf{R}^{(1)} \right )^\mathrm{T}}{T} & \frac{\mathbf{R}^{(2)} \left ( \mathbf{R}^{(2)} \right )^\mathrm{T}}{T}
\end{array} \right ).
\end{equation}
So, the asymmetric correlation matrix (for $\tau = 0$) $\mathbf{k} = \mathbf{R}^{(1)} ( \mathbf{R}^{(2)} )^\mathrm{T} / T$ and its transpose are embedded as the off-diagonal blocks of a larger object, which we shall call joint correlation matrix, having real eigenvalues. 

\begin{figure}
\begin{center}
\resizebox{0.99\columnwidth}{!}{
  \includegraphics{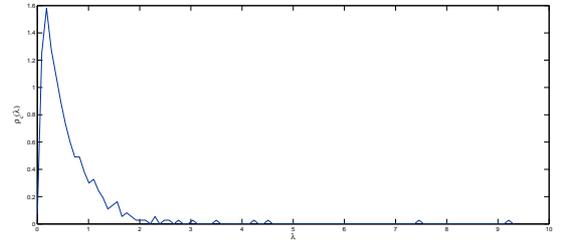}}
\end{center}
\caption{Eigenvalue spectrum of the joint correlation matrix in equation \eqref{jointmatr}. For better visualization, the two largest eigenvalues, equal to 112.7 and 31.8, have not been plotted.}
\label{BulkC}
\end{figure}

The eigenvalue spectrum of the joint correlation matrix in equation \eqref{jointmatr} displays one main bulk 
(see Figure \ref{BulkC}), plus a few eigenvalues ``leaking out'' of such bulk. Some of those can already be seen in Figure \ref{BulkC}, but not the largest two, equal to $\lambda_1 = 112.7$ and $\lambda_2 = 31.8$, 
\emph{i.e.} much larger than all the remaining ones. Very interestingly, some intuition on the meaning of such eigenvalues can be grasped by means of principal component analysis. Let us denote the eigenvalues of the joint correlation matrix $\mathbf{c}$ as $\lambda_1 > \lambda_2 > \ldots > \lambda_{2N}$, and the corresponding normalized eigenvectors as $\mathbf{V}^{(i)} = (V_1^{(i)}, \ldots, V_{2N}^{(i)})$, for $i = 1, \ldots, 2N$. Denoting principal components as $e_i$, equation \eqref{PCdec} can be specialized to the present case by writing

\begin{equation} \label{PCdec2}
R_{it} = \sum_{j=1}^{2N} \sqrt{\lambda_j} \ V_i^{(j)} e_{jt}.
\end{equation}
In the above equation, values of the index $i$ going from $1$ to $N$ cover stocks belonging to the S$\&$P Index, while values going from $N+1$ to $2N$ refer to stocks in the FTSE Index. Given the above considerations on the eigenvalue spectrum of the $\mathbf{c}$ matrix, it is certainly interesting to look at the eigenvector components of $\mathbf{V}^{(1)}$ and $\mathbf{V}^{(2)}$, \emph{i.e.} the eigenvectors 
\begin{figure}
\begin{center}
\resizebox{0.99\columnwidth}{!}{
  \includegraphics{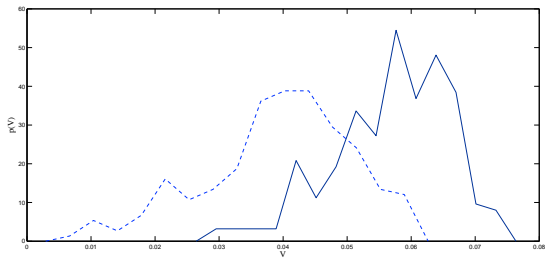}}
\end{center}
\caption{Component distribution for the eigenvector $\mathbf{V}^{(1)}$ related to the largest eigenvalue $\lambda_1$ of the joint correlation matrix in equation \eqref{jointmatr}. The distribution of components related to S$\&$P stocks is plotted with the solid line, while the one of components related to FTSE stocks is plotted with the dashed line.}
\label{EigDistr1}
\end{figure}
\begin{figure}
\begin{center}
\resizebox{0.99\columnwidth}{!}{
  \includegraphics{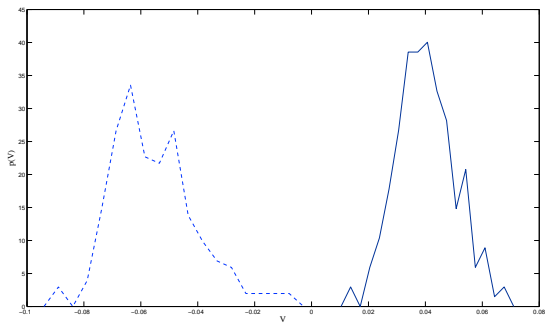}}
\end{center}
\caption{Component distribution for the eigenvector $\mathbf{V}^{(2)}$ related to the second largest eigenvalue $\lambda_2$ of the joint correlation matrix in equation \eqref{jointmatr}. The distribution of components related to S$\&$P stocks is plotted with the solid line, while the one of components related to FTSE stocks is plotted with the dashed line.}
\label{EigDistr2}
\end{figure}
corresponding to the largest eigenvalues. In Figure \ref{EigDistr1} the components of $\mathbf{V}^{(1)}$ are reported, distinguishing those related to S$\&$P stocks (solid line) from those related to FTSE stocks (dashed line). As one can see, both component groups are positive and they partially overlap. Thus, from equation \eqref{PCdec2} one can conclude that the first principal component of the $\mathbf{c}$ approximately impacts all stocks in the same way. On the contrary, one can see in Figure \ref{EigDistr2} that the eigenvector components of $\mathbf{V}^{(2)}$ are split into two well separated groups: components related to S$\&$P stocks are positive, while component related to FTSE stocks are negative. Also, one can
\begin{figure}
\begin{center}
\resizebox{0.99\columnwidth}{!}{
  \includegraphics{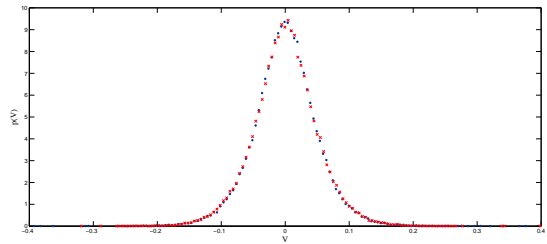}}
\end{center}
\caption{Component distribution of the eigenvectors $\mathbf{V}^{(i)}$ of the joint correlation matrix in equation \eqref{jointmatr} for $i = 3, \ldots, 2N$. Blue dots refer to components related to S$\&$P stocks, whereas red crosses refer to to components related to FTSE stocks.}
\label{EigDistrRest}
\end{figure} 
verify that the component distributions for all the remaining eigenvectors (from $\mathbf{V}^{(3)}$ to $\mathbf{V}^{(2N)}$) almost exactly overlap. These facts suggest the following interpretation. The largest eigenvalue $\lambda_1$ is a ``global market'' eigenvalue, meaning that the corresponding principal component, accounting for $28.2\%$ of the overall data variance, drives both markets in the same direction (all $V_i^{(1)}$s positive), and roughly drives all of their stocks with the same intensity (partial overlap of the two distributions in Figure \ref{EigDistr1}). On the other hand, Figure \ref{EigDistr2} makes it clear that the principal component related to the second largest eigenvalue, accounting for almost $8\%$ of the overall data variance, is the main source of negative correlation between the two markets under study. Those observations essentially match the results discussed in Section \ref{PCmap}. In particular, in Figure \ref{K11} it was shown that the main principal components of the two markets are strongly correlated for $\tau = 0$, whereas their second most relevant principal components are negatively correlated. So, both analyses point out two main sources of correlation, one positive and one negative, between the two markets. The remaining eigenvalues of $\mathbf{c}$ do not allow for similarly clear interpretations, and this is quite well portrayed by the almost overlapping eigenvector component distributions shown in Figure \ref{EigDistrRest}.

\section{Conclusions}
\label{concl}
In very general terms, the main motivation for the study presented in this paper was to look for an empirical realization of a random asymmetric generalized correlation matrix of the type \eqref{Pearson12tau} and its eigenvalue density (equations \eqref{Kdensity} and \eqref{Keffdensity}), attempting to perform a correlation analysis with complex eigenvalues. Financial data were chosen as a case study, but all of the analyses performed could be exactly replicated in any context where time series are involved.

As already stated, looking at eigenvalues might represent a limitation, since it forces one to work with square matrices. From the financial viewpoint, this limitation forced us to work with equal number of stocks in the
S$\&$P and FTSE datasets. Drawing more significant conclusions on the possible correlations between the two whole indices (or markets) would require to keep the datasets to their actual dimensions, and consequently to work with singular values, as in \cite{Bouchaud07}. Still, whenever one is reasonably allowed to work with an approximately close number of variables in the two sub-systems, the radial density \eqref{Keffdensity} represents an effective tool to detect the presence of cross-correlations (as in Figure \ref{Bulk}) or autocorrelations almost at first glance, or at least after a quick fitting procedure to determine the effective value of the $q$ ratio.

As far as financial aspects are concerned, all the results we presented suggest that all macroscopically relevant correlations between the New York and London stock exchanges expire within a 24 hours time window. Switching to principal components and studying the spectral properties of the joint correlation matrix \eqref{jointmatr} allowed us both to corroborate such findings and to unravel some other non trivial facts, such as the identification of the main sources of positive and negative correlation between the markets we considered, and the emergence of an effective system dimensionality due to autocorrelations in the principal components. Also, it would definitely be interesting to repeat all or some of the analyses detailed in this paper on high frequency data, possibly comparing the results with those presented in the previously mentioned paper \cite{Drozdz06}.

Lastly, from the viewpoint of RMT, equation \eqref{corrlink2} represents a very interesting starting point for possible future developments. More specifically: principal component analysis grants us that the variables which give rise to the $\mathbf{k}^{(e)}(\tau)$ matrix are exactly uncorrelated within each sub-system. So, whenever those are reasonably well described by Gaussian statistics, we know that the average eigenvalue density of $\mathbf{k}^{(e)}(\tau)$ is given by equations \eqref{Kdensity} and \eqref{Keffdensity} (possibly for some effective value of $q$, as we discussed). Thus, equation \eqref{corrlink2} describes the transition from the eigenvalue density arising from two uncorrelated systems (encoded in $\mathbf{k}^{(e)}(\tau)$) to the one of two systems having the correlation structure encoded in the $\mathbf{W}^{(1)}$ and $\mathbf{W}^{(2)}$ matrices (see equation \eqref{Wmatr}). This is an interesting property at least for the following reason. As far as theoretical advances in RMT are concerned, one could try to use recently developed tools about the multiplicative structure of random matrices \cite{Burda11_2} in order to derive analytical, or semi-analytical, results for the spectrum of the $\mathbf{k}(\tau)$ matrix seen as the outcome of the multiplicative action of two fixed known matrices ($\mathbf{W}^{(1)}$ and $\mathbf{W}^{(2)}$) on a known spectrum (the one given by $\mathbf{k}^{(e)}(\tau)$). Also, generalizing the results in equations \eqref{Kdensity} and \eqref{Keffdensity} to the eigenvalue spectra of asymmetric correlation matrices arising from random variables displaying both cross-correlations and autocorrelations would represent a major challenge to RMT developments. However, intuition based on similar generalizations for ordinary correlation matrices (see for example \cite{Burda05}) suggests that the presence of short lived, \emph{e.g.} exponentially damped, autocorrelations would not modify the eigenvalue spectra in a dramatic fashion. \\

\noindent We thank Guido Montagna for helpful suggestions and for reading the preliminary version of our manuscript. G. L. also wishes to thank Oreste Nicrosini and Andrea Schirru for many stimulating discussions during the early stages of this work.

%


\end{document}